\documentclass[fleqn,twoside]{article}
\usepackage{espcrc2}
\usepackage{epic}

\usepackage[dvips]{graphicx}

\newcommand{\N}{\mathcal{N}}
\newcommand{\der}{\partial}
\newcommand{\unity}{\mbox{\unboldmath$\lefteqn{\mathsf{1}}%
\hspace*{0.12em}\mathsf{1}$}}
\newcommand{\ddx}[1]{\frac{\partial}{\partial{#1}}}
\newcommand{\Qt}{\tilde{Q}}
\newcommand{\thetat}{\tilde{\theta}}
\newcommand{\varphit}{\tilde{\varphi}}
\newcommand{\phit}{\tilde{\phi}}
\newcommand{\chit}{\tilde{\chi}}
\newcommand{\hn}{\hat{n}}
\newcommand{\ha}{\hat{a}}
\newcommand{\hta}{\hat{\tilde{a}}}
\newcommand{\Psibar}{\bar{\Psi}}

\newcommand{\cbar}{\bar{c}}
\newcommand{\xibar}{\bar{\xi}}
 \newlength{\figwidth}

\title{Twisted $\N=2$ exact SUSY on the lattice for BF and
Wess-Zumino\thanks{Talk presented by I. Kanamori.}\thanks{This talk is
supported by the 2004 Hokkaido University International Exchange Program
Fund.}}

\author{Alessandro D'Adda\address{INFN sezione deTorino, and
Dipartimento di Fisica Teorica, Universita di Torino, 
I-10125 Torino, Italy}, 
Issaku Kanamori\address[H]{Department of Physics, Hokkaido University,
Sapporo, 060-0810, Japan}, Noboru Kawamoto\addressmark[H] and Kazuhiro
Nagata\addressmark[H]}
       
\begin{document}

\begin{abstract}
We formulate exact supersymmetric models on a lattice.
We introduce noncommutativity to ensure the Leibniz rule.
With the help of superspace formalism, we give supertransformations
which keep the $\N=2$ twisted SUSY algebra exactly.
The action is given as a product of (anti)chiral superfields on the
lattice.
We present BF and Wess-Zumino models as explicit examples of our
formulation. Both models have exact $\N=2$ twisted SUSY in 2 dimensional
space at a finite lattice spacing.
In component fields, the action has supercharge exact form.

\vspace{1pc}
\end{abstract}

\maketitle

\section{Introduction}
In order to formulate supersymmetric theories nonpertubatively, 
a lattice formulation is required in which supersymmetry(SUSY) is
exactly reserved. 
It is well-known, however, that there are some difficulties to keep the
SUSY exact. One of the difficulties comes from the breakdown of the
Leibniz rule. 
Since the SUSY algebra contains derivatives, its breakdown is crucial.

Keeping this point in mind, we adapt following approach \cite{ourwork}:
\begin{itemize}
 \item Discretization of SUSY algebra itself
 \item Twisted SUSY
 \item Superspace with ``mild noncommutativity''
\end{itemize}
The discretization breaks the Leibniz rule, but the noncommutativity
restores it.  We construct supersymmetric abelian BF and Wess-Zumino
model in 2 dimensions as explicit examples of our approach.
In this talk we mainly concentrate on free cases.
The basic idea of our formulation  is given by
the talk of N.~Kawamoto,
some other models with interactions in the poster of K.~Nagata \cite{proc}.
The continuum version of our model can be found in \cite{KKU}.

\section{Twisted Supersymmetry}
$\N=2$ twisted SUSY algebra \cite{Witten} is,
\begin{eqnarray}
 \lefteqn{\{Q, Q_\mu\} = i\der_\mu, \qquad
   \{\Qt, Q_\mu\}=-i\epsilon_{\mu\nu}\der^\nu,}\\
 \lefteqn{ Q^2 = \Qt^2= \{Q,\Qt\}=\{Q_\mu,Q_\nu\}=0, }
\end{eqnarray}
where $Q$'s are twisted supercharges and related to usual supercharges
of Majorana spinor by
\begin{equation}
  Q_{\alpha i}
    = \bigl( \unity Q + \gamma_\mu Q_\mu + \gamma_5 \Qt\bigr)_{\alpha i}.
\end{equation}
The $\gamma$ matrices are given by $\gamma_1 = \sigma_3$, $\gamma_2 =
\sigma_1$ and $ \gamma_5 = \gamma_1\gamma_2$.
The original charge $Q_{\alpha i}$ has two kinds of suffices,
spinor($\alpha$) and extended SUSY($i$). 
We regard $i$ as a spinor suffix and mix both suffices through the twist.

The twisted supercharges are no longer spinors. $Q$ is a scalar, $Q_\mu$
a vector and $\Qt$ a pseudo scalar. These quantities are much easier to
treat on the lattice than spinors.  This is an advantage of
twisted SUSY.  
Another advantage is that the action is given by supercharge-exact form so 
the invariance is manifest because of the nilpotency of $Q$'s.
Some authors use this advantage to formulate twisted SUSY on lattice
\cite{catterall,sugino}.

\section{Superspace with Noncommutativity}
Next we discretise the algebra and show how the ``mild noncommutative''
superspace arises.

Consider for instance a commutator:
\begin{equation}
 [\theta Q, \theta_\mu Q_\mu] = i\theta\theta_\mu \der_{+\mu},
 \label{eq:discretized}
\end{equation}
where we use Grassmann parameters $\theta$ and $\theta_\mu$, and
a forward difference $\der_{+\mu}f(x) = f(x+2\hn_\mu) - f(x)$ as a derivative.

Operating the r.h.s of eq.(\ref{eq:discretized}) on a product of functions,
\begin{eqnarray}
 \theta\theta_\mu \der_{+\mu} (f(x)g(x))
  &\!\!\!\!=\!\!\!\!&
   \bigl( \theta\theta_\mu \der_{+\mu} f(x)\bigr)g(x) \nonumber\\*
  && \hspace{-1.5em} {}+ \theta\theta_\mu f(x+2\hn_\mu) \der_{+\mu}g(x),
\end{eqnarray}
we obtain the breakdown of Leibniz rule, 
where the argument of $f$ is shifted.
Introducing following ``mild noncommutativity'',
\begin{equation}
 \theta\theta_\mu f(x+2\hn_\mu) = f(x)\theta\theta_\mu,
\end{equation}
we can recover the usual Leibniz rule.
Thus in order to compensate the breakdown of Leibniz rule, we 
need noncommutative Grassmann parameters.
And these parameters naturally lead us to superspace formulation.
Note this kind of noncommutativity is discussed in the context of
differential forms on the lattice and gauge theories \cite{dimakis,KK}.

On the superspace, supercharges are given by the following derivative
operators:
\begin{eqnarray}
  Q &=&\ddx{\theta}+\frac{i}{2}\theta \der_{+\mu}
  , \quad
  \Qt = \ddx{\thetat} - \frac{i}{2}\theta^\mu\epsilon_{\mu\nu}\der^{-\nu},
 \nonumber\\
 Q_\mu &=& \ddx{\theta^\mu}+\frac{i}{2}\theta\der_{+\mu}
        -\frac{i}{2}\thetat\epsilon_{\mu\nu}\der_{-\nu},
\label{eq:charges}
\end{eqnarray}
which satisfy
\begin{eqnarray}
 \{Q,Q_\mu \} &\!\!=\!\!& i\der_{+\mu}
  ,\qquad
  \{ \tilde{Q}, Q_\mu \} = -i\epsilon_{\mu\nu}\der_{-\nu},
 \\
 \{Q,\tilde{Q}\} &\!\!=\!\!& \{Q_\mu, Q_\nu\}=Q^2 = \tilde{Q}^2 = Q_\mu^2 = 0.
\end{eqnarray}
$Q_A$ has a shifting nature of $Q_A f(x) = f(x+2\ha_A)Q_A$ for $Q_A
= \{Q, Q_\mu, \Qt\}$.  We cannot determine parameter $\ha_A$ uniquely
but in this talk we use the following symmetric choice:
\begin{eqnarray}
 2\ha &=& - 2\hta = \hn_1+\hn_2,\\
 2\ha_1 &=& -2\ha_2 = \hn_1-\hn_2.
\end{eqnarray}
See Kawamoto's talk and Nagata's poster to more general choice and
details.
Here we only point out that 
we need both forward and backward difference operator $\der_{\pm\mu}$
to obtain nontrivial $\ha_A$.

We can define chiral($\Psi$) and anti-chiral($\Psibar$) 
superfield using our noncommutative $\theta_A$:
\begin{eqnarray}
  \Psi(x) 
   &=& ic(x) + \theta^\mu \omega_\mu(x+\ha_\mu)\nonumber\\*
   &&  {}     + i\theta^1\theta^2 \lambda(x) + \cdots,\\
  \Psibar(x)
  &=& i\cbar(x) + \theta b(x+\ha) + \tilde{\theta}\phi(x+\hta) \nonumber\\*
  &&{}    -i \theta\tilde{\theta} \rho(x)+\cdots.
\end{eqnarray}
Here ``$\cdots$'' contains only derivative terms.
The component of $\theta^A \theta^B \ldots$ lives 
on the site of $x+\ha_A + \ha_B +\cdots$ due to the
noncommutativity of $\theta_A$.
Thus the superfield has semi-local structure which goes to local in the
naive continuum limit because $\ha_A$ vanishes in this limit. 
The transformation of each component field can be read from using
eq.(\ref{eq:charges}):
\begin{equation}
 Q_A \Psi (x) =s_A\Psi(x), \quad Q_A \Psibar(x) = s_A\Psibar(x).
\end{equation}
We use $s_A$ as transformations of the components and distinguish from $Q_A$.

\section{Actions}
Having defined almost the same tools as in the continuum, 
we can straightforwardly construct actions on the lattice.

The most simple one is a product of chiral and anti-chiral superfields:
\begin{eqnarray}
  S
   &=& \sum_x \Psibar(x) \Psi(x)\Bigm|_{\theta^4}\\
   &=& \sum_x i s_1 s_2 \tilde{s} s \bigl(\cbar(x)c(x)\bigr)
       \label{eq:BF-s-exact}\\
   &=& \sum_x\Bigl[
        \phi(x+a) \epsilon_{\mu\nu}\der_{+\mu}\omega_\nu(x+\ha_\nu)
  \nonumber\\*
   && {}+ b(x-\ha) \der_{-\mu}\omega_\mu(x+\ha_\mu)
  \nonumber\\*
   && {}-i \cbar(x) \der_{+\mu}\der^{\!-\mu} c(x)
        + i\rho(x)\lambda(x)
	  \Bigr],
 \label{eq:BF}
\end{eqnarray}
where superfield $\Psi$ and $\Psibar$(and their lowest components
$c$ and $\cbar$)
are fermionic.
This action is supersymmetric BF with gauge fixing term,
ghost($c, \cbar$) and fermionic auxiliary fields($\lambda, \rho$).
Note it is supercharge exact as in eq.(\ref{eq:BF-s-exact}), and all
supercharges are nilpotent, thus the SUSY invariance is manifest for all
the four charges.

The action (\ref{eq:BF}) has a kind of dual structure with the
symmetric choice of $\ha_A$.
Since we adapted double size convention for difference operator, 
we first set $x=(\mbox{even},\mbox{even})$.  
Then the noncommutativity shifts the argument
by $2\ha_A$, for example,
\begin{equation}
 s(\cbar(x)c(x)) = (s \cbar(x))c(x)  - \cbar(x+2\ha)sc(x),
\end{equation}
where $x+2\ha_A = (\mbox{odd},\mbox{odd})$. 
Thus we need $c$ on both $(\mbox{even},\mbox{even})$ and
$(\mbox{odd},\mbox{odd})$. 
And bosons live on the dual site while fermions on the
original(Fig.~\ref{fig:boson-fermion}).
\begin{figure}
 \begin{center}
  \setlength{\figwidth}{.7\linewidth}
  \input{boson-fermion.tex}
 \end{center} 
\vspace{-2.5\baselineskip}
 \caption{Fermions($\bullet$) and bosons($\times$) in BF case.
$c$ lives on $\bullet$.}
 \label{fig:boson-fermion}
\end{figure}

Next, let us change statistics of the superfields. 
Using bosonic($\varphi, \phi, \varphit, \phit$) and
fermionic($\chi, \chit, \psi_\mu$) components,
we obtain
\begin{eqnarray}
  S  &\!\!\!=\!\!\!& \sum_x\Bigl[
     \varphi(x)\der^{+\mu}\der_{-\mu}\phi(x) + \varphit(x)\phit(x)\nonumber\\
    && \hspace*{-3em}
   {}-i\bigl(\der_{+\mu}\chi(x-\ha)
        -\epsilon_{\mu\nu}\der^{-\nu}\chit(x+\ha)\bigr)
       \psi^\mu(x+\ha_\mu)\Bigr].\nonumber\\
\end{eqnarray}
After untwisting the fermions
\begin{eqnarray}
 \xi_{\alpha i}(x)
  &=&  \frac{1}{2}\bigl( \unity\chi(x-\ha) + \gamma_\mu\psi^\mu(x+\ha_\mu)
      \nonumber\\
   && \hspace*{2em}{}
        + \gamma_5\chit(x+\ha) \bigr)_{\alpha i},
\end{eqnarray}
and suitable redefinition of bosons, we obtain $\N=2$ Wess-Zumino model:
\begin{eqnarray}
 S  &\!\!\!=\!\!\!& \sum_x \Bigl[
   -\phi_i(x) \der^{+\mu}\der_{-\mu}\phi_i(x)
   +F_i(x) F_i(x)\nonumber\\
 &&\hspace*{-2em}
    {}+\frac{i}{2}\xibar_{i\alpha}(x)(\gamma_\mu)_{\alpha\beta}
     (\der_{+\mu}+\der_{-\mu})\xi_{\beta i}(x) \\
 &&\hspace*{-2em}
    {}-\frac{i}{2}\xibar_{i\alpha}(x)(\gamma_5)_{\alpha\beta}
     (\der_{+\mu}-\der_{-\mu})\xi_{\beta j}(x)
    (\gamma_5\gamma_\mu)_{ji}
    \Bigr]. \nonumber
\end{eqnarray}
The fermion part is nothing but staggered fermion in K-S form.
Note that the flavor(or taste) index of staggered fermion $i$ now becomes 
that of extended SUSY.
We call this relation ``Dirac-K\"ahler mechanism'' since staggered 
fermion is a lattice version of Dirac-K\"ahler fermion. 
See Fig.~\ref{fig:D-K}. 
We would like to comment that a different use of staggered fermion is found
in \cite{ichimatsu}.
\begin{figure}
 \begin{center}
  \setlength{\figwidth}{.7\linewidth}
  \input{D-Ksimple.tex}
 \end{center} 
\vspace{-2.5\baselineskip}
 \caption{$\chi$, $\psi_\mu$ and $\chit$ correspond to $0$, $1$ and
 $2$-form on the thick lattice, which make Dirac-K\"ahler fermion $\xi$.}
 \label{fig:D-K}
\end{figure}

\section{Discussion}
We have defined lattice models which keep exact $\N=2$ SUSY for all
charges.  
We use mild noncommutativity to compensate the breakdown
of Leibniz rule.
Since supercharges also have noncommutativity, our
definition of SUSY may be a new one in the same sense that Ginsparg-Wilson
relation defines a new chiral symmetry.

\end{document}